\documentclass[useAMS,usenatbib]{mn2e}
\usepackage{amsmath,amssymb}
\usepackage{graphicx}
\title[Supermassive Black Holes]{Supermassive Black Holes: Connecting the Growth to the Cosmic Star Formation Rate}
\author[E.S. Pereira and O. D. Miranda]{Eduardo S. Pereira\thanks{duducosmo@das.inpe.br} and Oswaldo D. Miranda\thanks{oswaldo@das.inpe.br}\\
INPE - Instituto Nacional de Pesquisas Espaciais - Divis\~{a}o de Astrof\'{i}sica,\\
Av. dos Astronautas 1758, S\~{a}o Jos\'{e} dos Campos, 12227-010 SP, Brazil}

\begin{document}

\date{\today}

\pagerange{\pageref{firstpage}--\pageref{lastpage}} \pubyear{2011}

\maketitle

\label{firstpage}

\begin{abstract}
In this Letter, we present a model connecting the cosmic star formation rate (CSFR) to the growth of supermassive black holes.
Considering that the evolution of the massive black hole is dominated by accretion (Soltan's argument) and that the accretion
process can be described by a probabilistic function directly regulated by the CSFR, we obtain the evolution of the black hole mass
density. Then using the quasar luminosity function, we determine both the functional form of the radiative efficiency
and the evolution of the quasar duty-cycle as functions of the redshift. We analyze four different CSFRs showing that the quasar
duty-cycle, $\delta(z)$, peaks at $z\sim 8.5-11$ and so within the window associated with the reionization of the Universe. In
particular, $\delta_{\rm max}\sim 0.09-0.22$ depending on the CSFR. The mean radiative efficiency, $\bar\eta(z)$, peaks at
$z\sim 0.1-1.3$ with $\bar\eta_{\rm max}\sim 0.10-0.46$ depending on the specific CSFR used. Our results also show that is not necessary
a supercritical Eddington accretion regime to produce the growth of the black hole seeds. The present scenario is consistent with the
formation of black hole seeds $\sim 10^3{\rm M}_\odot$ at $z\sim 20$.
\end{abstract}


\section{Introduction}\label{sec:intro}

There is strong evidence that nearly all galaxies contain supermassive black holes (SMBHs) in their centers, and that the evolution of the supermassive
black hole and its host galaxy are connected (see, e.g., \citealp{ferrarese00}). In particular, the masses of the black holes range from $\sim 10^{6}{\rm M}_{\odot}$ for
galaxies with small bulges up to $\sim 10^{9}{\rm M}_{\odot}$ for galaxies in cores of groups and clusters of galaxies (see, e.g., \citealp{margo98}).
These SMBHs are present not only in the local Universe, in the form of the low luminosity active galactic nuclei (AGN), but also in the early stages of galaxy
formation as can be seen from quasars discovered beyond $z > 6$ (see, e.g., \citealp{fan03}). Furthermore, accretion onto massive black holes is generally accepted
as a way to power strong emission as observed in the cases of AGNs and quasars.

On the other hand, it is reasonable to consider that the growth of SMBHs can be regulated, in some way, by the cosmic star formation
rate - CSFR (see, e.g., \citealp{franceschine99,haiman04,heckman04,merloni04,mahmood05,wang06}).
This is reinforced by the fact that the CSFR can be directly connected to the masses of dark matter halos (see, e.g.,
\citealp{conroy09,pereira10}). The dark halos are the natural nursery for the birth and growth of the massive black holes. In
principle, measuring the masses and accretion rate of the black holes that drive the AGNs and quasars could help us understand
the evolution of these sources, their connection with the CSFR, and the contribution of mini-quasars to the reionization of the
universe. Here, we present a formalism permitting to confront several CSFRs with the quasar luminosity function (QLF). This formalism
could also be used for a better estimate of the CSFR up to redshift $\sim 7$ using the QLF as an observational data to be fitted by
the `theoretical' CSFRs. Furthermore, this work could contribute to the study of the feedback processes associated with both: star
formation at higher redshifts and growth of supermassive black holes. As main results, we derive the functional form of the radiative
efficiency associate to the accretion process of these black holes and the evolution of the quasar duty-cycle. Through this paper we
consider standard cosmological model ($\Lambda$CDM) with $\Omega_{\rm b} = 0.04$, $\Omega_{\rm m} = 0.27$, $\Omega_{\Lambda} = 0.73$,
$h = 0.73$.

\section{SMBHs and the CSFR}\label{sec:1}

We consider that black holes grow by accreting matter (Soltan's argument; see, e.g., \citealp{soltan82,wang08}) and that the accretion process can be
described by a probabilistic function directly regulated by the CSFR. This can be described in the following way

\begin{equation}\label{mmass}
\rho_{\rm BH}(z) = \rho^ {0}_{\rm BH} \frac{{\rho}_{{\star}{\rm BH}}(z)}{{\rho}_{{\star}{\rm BH}}(z=0)},
\end{equation}

\noindent where $\rho^{0}_{\rm BH}$ represents the black hole mass density in our local universe. The function
${\rho}_{{\star}{\rm BH}}(z)$ makes the connection between the CSFR and the black hole mass density at redshift $z$.
In particular, we consider

\begin{equation}\label{cbhda}
{\rho}_{\star \rm BH}(z)=\int^{z_{f}}_{z}{\frac{\dot{\rho}_{\star}(z')}{(1+z')}P(t_{\rm d})}\frac{dt_{\rm d}}{dz'}dz'.
\end{equation}

In Eq. (\ref{cbhda}), $\dot{\rho}_{\star}(z)$ represents the CSFR at redshift $z$, $P(t_{\rm d})$ is the probability per unit of time of the black hole
grow up by accreting matter from the environment, and the $(1+z)$ term in the denominator considers the time dilatation due to the cosmic expansion.
The time delay $t_{\rm d}$ makes the connection between the redshift $z_{\rm f}$ at which the accretion disk forms around the black hole and the redshift $z$
at which the material is incorporated into the black hole. Thus, we have
\begin{equation}
t_{\rm d}= \int^{z_{\rm f}}_{z}\frac{9.78\,h^{-1} \rm{Gyr}}{(1+z')\sqrt{\Omega_{\Lambda}+\Omega_{\rm m}(1+z')^{3}}}dz'.
\end{equation} 

We consider that $P(t_{\rm d})=Bt_{\rm d}^{n}$ and this function is normalized as $\int^{t_{\rm now}}_{t_{\rm min}} P(t_{\rm d})
dt_{\rm d}=1$, where $t_{\rm min}$ is the minimum time required for the accretion process to take place and
$t_{\rm now}$ is the age of the universe. We assume $t_{\rm min}=t_{\rm s}=4.2 \times 10^{7}\;{\rm yr}$, where $t_{\rm s}$ is the
Salpeter time. It is worth stressing that $P(t_{\rm d})\propto t_{\rm d}^{n}$ has been used by different authors in several
astrophysical contexts (see, e.g., \citealp{regimbau06,regimbau09} and references therein).

\begin{figure}
\includegraphics[width=94mm]{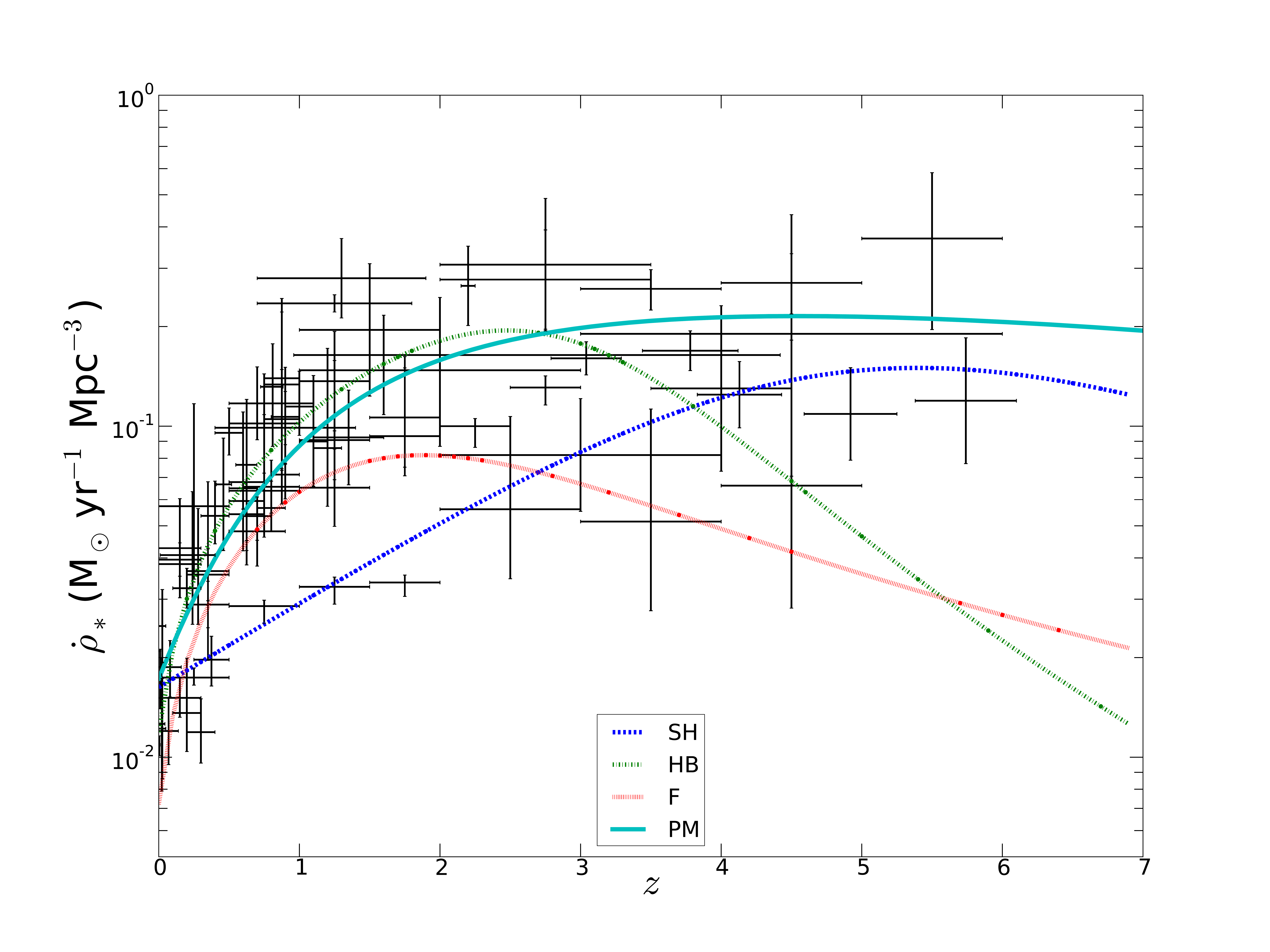}
\caption{Cosmic Star Formation Rate (CSFR) derived by: PM - \citet{pereira10}; SH - \citet{springel03};
HB - \citet{hopbea06}; F - \citet{fardal07}. The observational points were taken from \citet{hokins04a,hopkins07a}.}\label{fig:csfr}
\end{figure}

In this work, we consider the CSFR, $\dot{\rho}_{\star}(z)$, derived by: PM - \cite{pereira10}
who obtained this function from the hierarchical scenario for structure formation using a Press-Schechter-like formalism; SH -
\cite{springel03} who obtain the CSFR from hydrodynamic simulations; HB - \cite{hopbea06} and F - \cite{fardal07} who derived the
CSFR using the available observational data. In Fig. \ref{fig:csfr}, we summarize all these different CSFRs.

\begin{center}
\begin{figure}
\includegraphics[width=94mm]{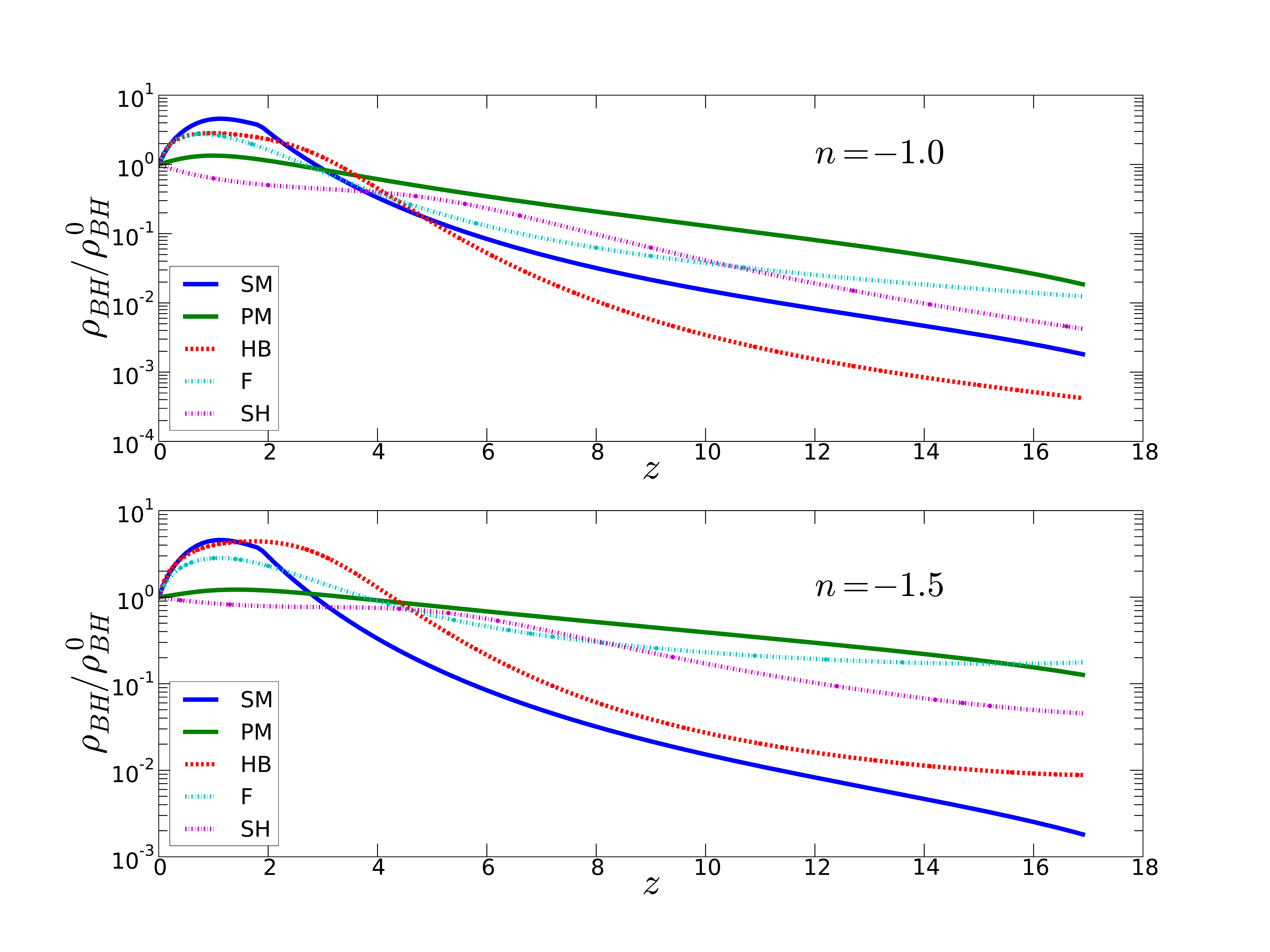}
\caption{The co-moving massive black hole mass density. We present $\rho_{\rm BH}(z)/\rho^{0}_{\rm BH}(z)$ using the four
CSFRs presented above. The results are shown for two values of the $n$-exponent of the probability function ($n=-1.0$ and $n=-1.5$).
We also present for comparison $\rho_{\rm BH}$ derived from the standard model (SM).}\label{fig2}
\end{figure}
\end{center}

In Fig. \ref{fig2} can be seen the black hole mass density as a function of the redshift. The results are presented for four different CSFRs, as explained above,
and using the formulation contained in Eqs. (\ref{mmass})-(\ref{cbhda}). Observe that for the CSFR derived by PM, $\rho_{\rm BH}/\rho^{0}_{\rm BH}$ peaks at
$z=1.1$ ($1.5$) for $n=-1.0$ ($n=-1.5$). Considering $n=-1.0$, note that PM-CSFR produces higher values for $\rho_{\rm BH}(z)$ at redshifts $z > 4$
than those produced by others CSFRs. We also plot in Fig. \ref{fig2} $\rho_{\rm BH}(z)$ derived from the standard model (SM). In
this case, $\rho_{\rm BH}(z)$ is obtained from the integration of the QLF, $\Phi_{\rm b}(L_{\rm b},z)$, in the following way
(see, e.g., \citealp{samll92,yutr02,merloni04,wang06b,shankar09b,shankar09})

\begin{equation}
\rho_{\rm BH}(z) = \int^{\infty}_{z}\frac{dt}{dz}\int^{L_{\rm s}}_{L_{\rm i}}{\frac{1-{\eta}}{c^{2}{\eta}}L_{\rm b}
\Phi_{\rm b}(L_{\rm b},z)}dL_{\rm b}\label{sbrho},
\end{equation}

\noindent where $L_{\rm b}$ is the bolometric luminosity taken from \cite{hopkins07}, ${\eta}$ is the radiative efficiency
and $c$ is the speed of light. Note that in order to solve equation (\ref{sbrho}) some consideration about ${\eta}$ value should be
made. In general, $\eta$ is assumed constant with typical value $\sim 0.1$. However, we will show in the next Section that ${\eta}$
must be a function of time if the growth of the SMBHs is regulated by the CSFR.

\section{The mean radiative efficiency and the quasar duty-cycle}\label{sec:2}

The bolometric luminosity of a black hole with accretion rate $\dot{M}_{\rm a}$ is given by $L= \bar{\eta}\dot{M}_{\rm a} c^{2}$ (with $\bar{\eta}$ the mean
radiative efficiency). Considering that $\dot{M}_{\rm a}$ is related to the mass variation of the black hole by $\dot{M}_{\rm a}=\dot{M}_{\rm BH}/(1.0-\bar{\eta})$,
and that $\bar{\eta}$ is a function only of the redshift $z$, we obtain:

\begin{equation}\label{lumidens}
{U}=c^{2}\frac{\bar{\eta}}{(1-\bar{\eta})}\dot{\rho}_{\rm BH},
\end{equation}

\noindent with ${U}$ being the luminosity density. Deriving the equation (\ref{mmass}) in $z$ and using the result in
(\ref{lumidens}) produces:

\begin{equation}\label{lumidens2}
{U}(z) = \frac{\bar{\eta}c^{2}}{1-\bar{\eta}}\frac{\rho^{0}_{\rm BH}}{\rho^{0}_{\star {\rm BH}}}\frac{\dot{\rho}_{\star}(z)}{(1+z)}P(t_{\rm d}).
\end{equation}

The temporal dependence of $\bar{\eta}$ reflects the fact that the active supermassive black holes have a finite lifetime
\citep{davislaor11}. In order to derive $\bar{\eta}(z)$, we shall use the luminosity density ${U}(z)$ obtained from the integration
of the QLF in the following way

\begin{equation}\label{dlumi}
{U}(z) = \int_{L^{\star}_{\rm b}}L_{\rm b}\Phi_{\rm b}(L_{\rm b},z)dL_{\rm b},
\end{equation}

\noindent where $L^ {\star}_{\rm b}$ is the lower limit of the QLF. Here, we take $L^{\star}_{\rm b}$ and $\Phi_{\rm b}(L_{\rm b},z)$ from \cite{hopkins07}.
Then, defining

\begin{equation}\label{feta}
f(z)\equiv \frac{\bar{\eta}}{1-\bar{\eta}},
\end{equation}

\noindent we can write

\begin{equation}\label{lumidens3}
{U}=c^{2}f(z)\dot{\rho}_{\rm BH}.
\end{equation}

Now, we define a functional  $f'(z,\vec{n_{\rm i}})$ which will be used to map $f(z)$ given by Eq. (\ref{feta}). In particular, $\vec{n_{\rm i}}$ is a vector
of parameters and so

\begin{equation}\label{fparam}
f'(z,b_{1},b_{2},t_{\rm q})=C_{0}\left[\left(\frac{t_{\rm u}(z)}{t_{\rm q}}\right)^{b_{1}}+\left(\frac{t_{\rm q}}{t_{\rm u}(z)}\right)^{b_{2}}\right]^{-1},
\end{equation}

\noindent where $C_{0}$ is a normalization constant which gives $\bar{\eta}(z=0)=\bar{\eta}^{0}$ (with $\bar{\eta}^{0}=0.1$, see,
e.g., \citealp{hopkins07}), $t_{\rm q}$ can be understood as a characteristic time-scale, $b_{\rm i}$ (${\rm i}=1,2$) are
dimensionless constants, and $t_{\rm u}(z)$ is the age of the Universe at redshift $z$.

The parametric form of Eq. (\ref{fparam}) is widely used in the literature. For example, Eq. (12) of \cite{hopkinsH09} and Eq. (13)
of \cite{hnl06} are similar to Eq. (\ref{fparam}) presented here. Furthermore, \cite{wang06} used the same parametric form
to determine the mass function of supermassive black holes. Note that, we are using the Eq. (\ref{dlumi}) as a source of information
to obtain $\bar{\eta}(z)$. Thus, using Eq. (\ref{fparam}) in Eq. (\ref{lumidens3}) is possible to write

\begin{equation}
e_{\rm i}(\vec{n_{\rm i}},{U}_{\rm i},z_{\rm i})=\Vert {U}(z_{\rm i}) -c^{2}f'(z_{\rm i},\vec{n_{\rm i}})\dot{\rho}_{\rm BH}(z_{\rm i})\Vert.
\end{equation}

The function that will be minimized is

\begin{equation}\label{obfun}
J(\vec{n_{\rm i}}) = \sum^{N-1}_{i=0}{e^{2}_{\rm i}(\vec{n_{\rm i}},{U}_{\rm i},z_{\rm i})}.
\end{equation}

See that $\bar{\eta}=f(z)/(1.0+f(z))$ and $f(z)=f'(z,\vec{n_{\rm i}}_{\rm best})$, where $\vec{n_{\rm i}}_{\rm best}$ gives the best fit from the  equation
(\ref{obfun}). In Table \ref{table:bfl} we present the best fit parameters which permit to derive the function $\bar\eta(z)$. 

\begin{center}
\begin{figure}
\includegraphics[width=90mm]{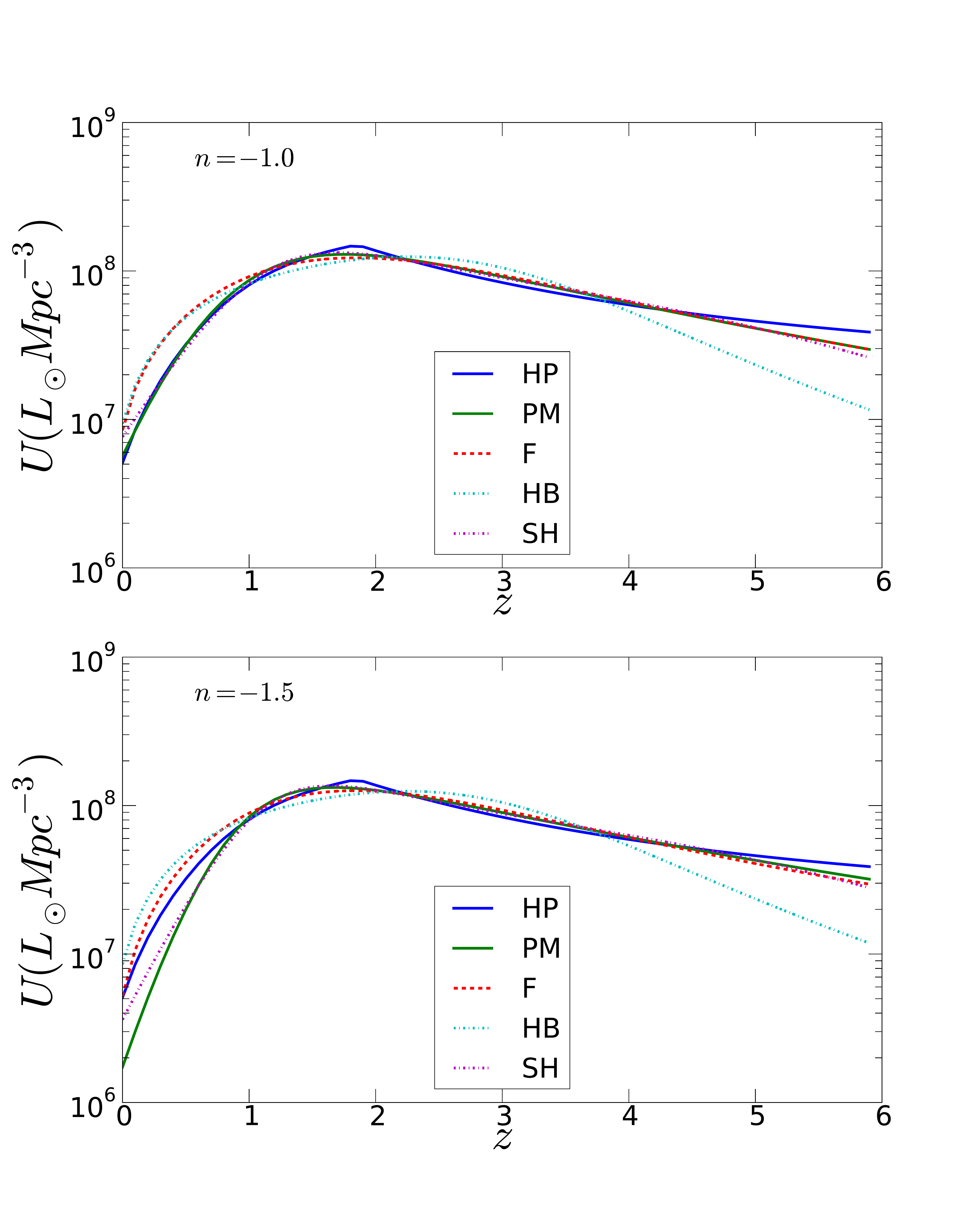}
\caption{The quasar luminosity density as a function of the redshift. HP represents $U(z)$ obtained from the integration of the QLF.
Different $U(z)$ directly derived from the CSFRs are also presented for $n=-1.0$ (upper panel) and $n=-1.5$ (lower panel).}
\label{fig3}
\end{figure}
\end{center}

Figure \ref{fig3} presents the luminosity density of this work when compared to that obtained from Eq. (\ref{dlumi}), which comes
from the integration of the QLF (\citealp{hopkins07}). In particular, for $n=-1.0$ we see that PM and SH CSFRs produce an excellent
agreement with $U(z)$ up to redshift $\sim 6$. In the case $n=-1.5$ we verify that F-CSFR has the best agreement with $U(z)$ at
$z\leq 1$. On the other hand, from $z\sim 1$ up to redshift $\sim 6$, PM and SH, beyond F-CSFR, have good agreement with the
integrated QLF if we consider the parametric form given by Eq. (\ref{fparam}).

\begin{table}
{\center
\caption{{Best fit parameters of $\bar{\eta}(z)$.}
\label{table:bfl}}
\begin{tabular}{@{}lcccc}
\hline
$n$ & CSFR & $b_{1}$ & $b_{2}$ & $t_{\rm q}(\rm Gyr)$\\
\hline 
-1.5 & PM & 2.85 & 2.39 & 5.74 \\
-1.5 & HB & 1.51 & 1.29 & 13.19 \\
-1.5 & F  & 0.82 & 1.47 & 5.99 \\
-1.5 & SH & 2.74 & 3.17 & 4.93 \\
-1.0 & PM & 1.81 & 1.96 & 5.40 \\
-1.0 & HB & 1.91 & 0.74 & 14.81 \\
-1.0 & F & 0.46 & 1.06 & 4.23 \\
-1.0 & SH & 2.37 & 2.72 & 4.72 \\
\hline
\end{tabular}

\medskip
}
\end{table}

\begin{center}
\begin{figure}
\includegraphics[width=94mm]{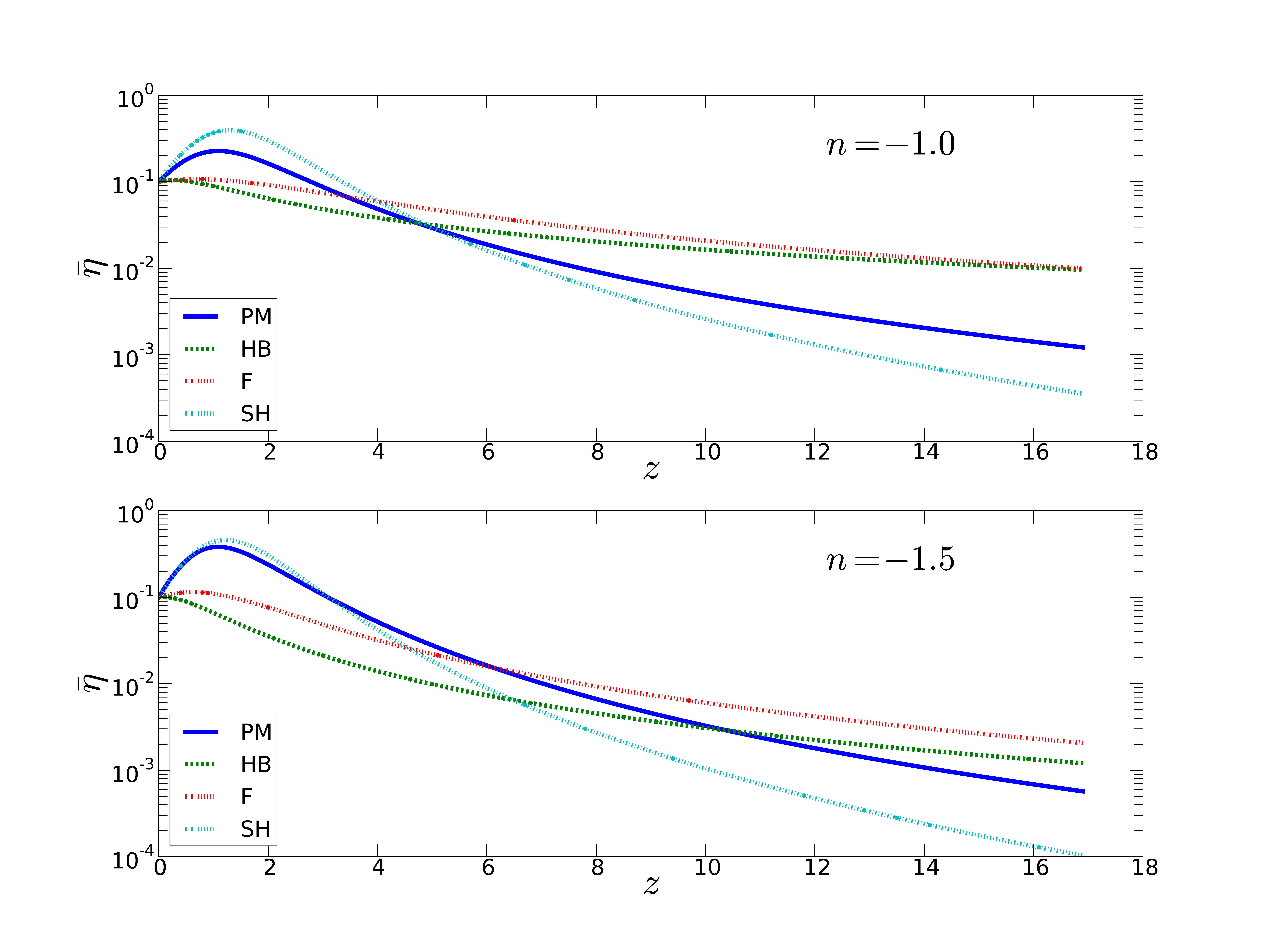}
\caption{Mean radiative efficiency as a function of the redshift. The upper panel presents the results for $n=-1.0$ while in the lower panel we have $n=-1.5$ for
the slope of the probability function.}
\label{fig:eta}
\end{figure}
\end{center}

In Fig. \ref{fig:eta} we present the mean radiative efficiency $\bar\eta(z)$ as a function of the redshift while in Table \ref{tbleta}
we show the redshift where $\bar\eta(z)$ peaks. See that $\bar\eta_{\rm max}$ is within the range $0.10-0.46$ depending on
the specific CSFR. In particular, in these cases we have $z_{\rm max}$ in the range $0.1-1.3$. These results are concordant with
accreting black holes which could reach $\bar{\eta} \sim 0.2$ for the most massive systems \citep{narayan05}. 

\begin{table}
{\center
\caption{{Maximum values for the mean radiative efficiency.}
\label{tbleta}}
\begin{tabular}{@{}lcccc}
\hline
$n$ & CSFR & $\bar\eta_{\rm max}$ & $z_{\rm max}$ \\
\hline
-1.5 & PM & 0.38 & 1.1 \\
-1.5 & HB & 0.10 & 0.1 \\
-1.5 & F  & 0.11 & 0.6 \\
-1.5 & SH & 0.46 & 1.2 \\
-1.0 & PM & 0.23 & 1.1 \\
-1.0 & HB & 0.10 & 0.3 \\
-1.0 & F  & 0.11 & 0.7 \\
-1.0 & SH & 0.39 & 1.3 \\
 \hline
\end{tabular}

\medskip
}
\end{table}

\begin{table}
{\center
\caption{{Maximum values for the quasar duty-cycle.}
\label{tblduty}}
\begin{tabular}{@{}lcccc}
\hline
$n$ & CSFR & $\delta_{\rm max}$ & $z_{\rm max}$ \\
\hline
-1.5 & PM & 0.13 & 11.1 \\
-1.5 & HB & 0.15 & 8.5 \\
-1.5 & F  & 0.09 & 9.9 \\
-1.5 & SH & 0.16 & 9.4 \\
-1.0 & PM & 0.19 & 10.9 \\
-1.0 & HB & 0.20 & 9.0 \\
-1.0 & F  & 0.14 & 9.7 \\
-1.0 & SH & 0.22 & 9.6 \\
 \hline
\end{tabular}

\medskip
}
\end{table}

On the other hand, the Eddington mass can be written as $\dot{M}_{\rm edd}={M_{\rm BH}}/{t_{\rm s}}$. The bolometric
luminosity-weighted by the Eddington mass is $L=(\eta c^{2} \dot{m} M_{\rm BH})/t_{\rm s}$, where  $\dot{m}=\dot{M}_{\rm a}
/\dot{M}_{\rm edd}$ is the dimensionless accretion rate. Assuming that $<\dot{m}>$ is a function only of the redshift
(see \citealp{hopkins07}) we can write

\begin{equation}\label{eq:5}
{U}(z)=\frac{\bar{\eta} <\dot{m}> \rho_{\rm BH} c^{2}}{t_{\rm s}}.
\end{equation}

Using equations (\ref{eq:5}) and (\ref{lumidens}), we obtain

\begin{equation}\label{eq:7}
<\dot{m}> =\frac{t_{\rm s}}{(1-\bar{\eta})}\frac{\dot{\rho}_{\rm BH}}{\rho_{\rm BH}}.
\end{equation}

\begin{center}
\begin{figure}
\includegraphics[width=94mm]{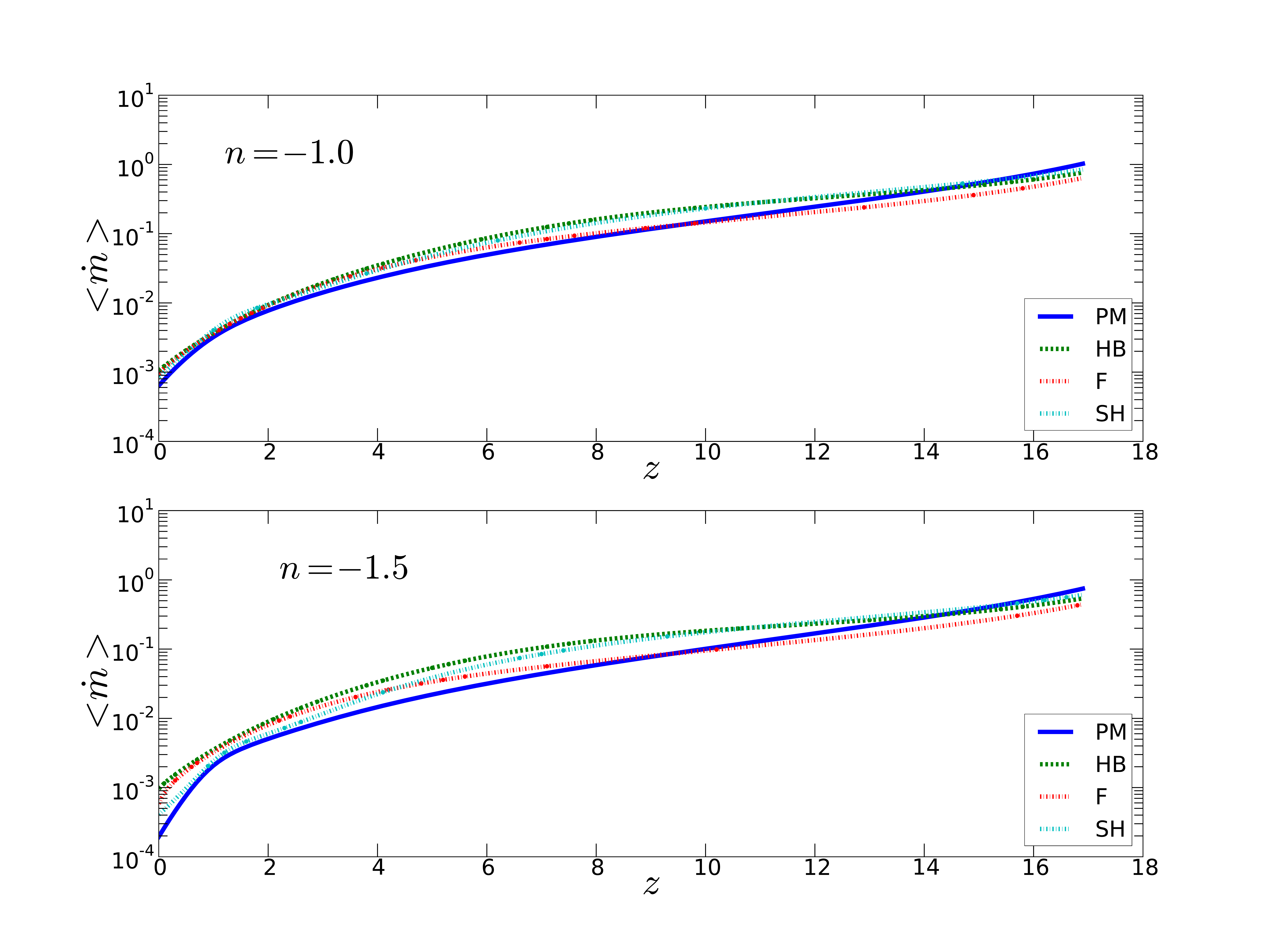}
\caption{The mean dimensionless accretion rate for $n=-1.0$ and $n=-1.5$ as functions of the redshift.}\label{fig:Mass}
\end{figure}
\end{center}

In Figure \ref{fig:Mass} we present the evolution of $<\dot{m}>$ with the redshift. We can see that the accretion processes
are more active at higher redshifts for all CSFRs studied in the present paper. However, our results also show that the accretion
processes never reach a super-Eddington regime. The accretion process associated to the growth of SMBHs is `well-behaved'
during all time associated from the seed black hole formation up to the present time. Another point can be derived from the equation
(\ref{eq:7}). It is possible to find the mean accretion-weighted by the lifetime of a SMBH \citep{merloni04}.
In particular, we have

\begin{equation}
\tau_{\rm DC}(z) = \int^{z_{\rm ini}}_{z}{<\dot{m}>\frac{dt}{dz'}dz'}.
\end{equation}

\begin{center}
\begin{figure}
\includegraphics[width=94mm]{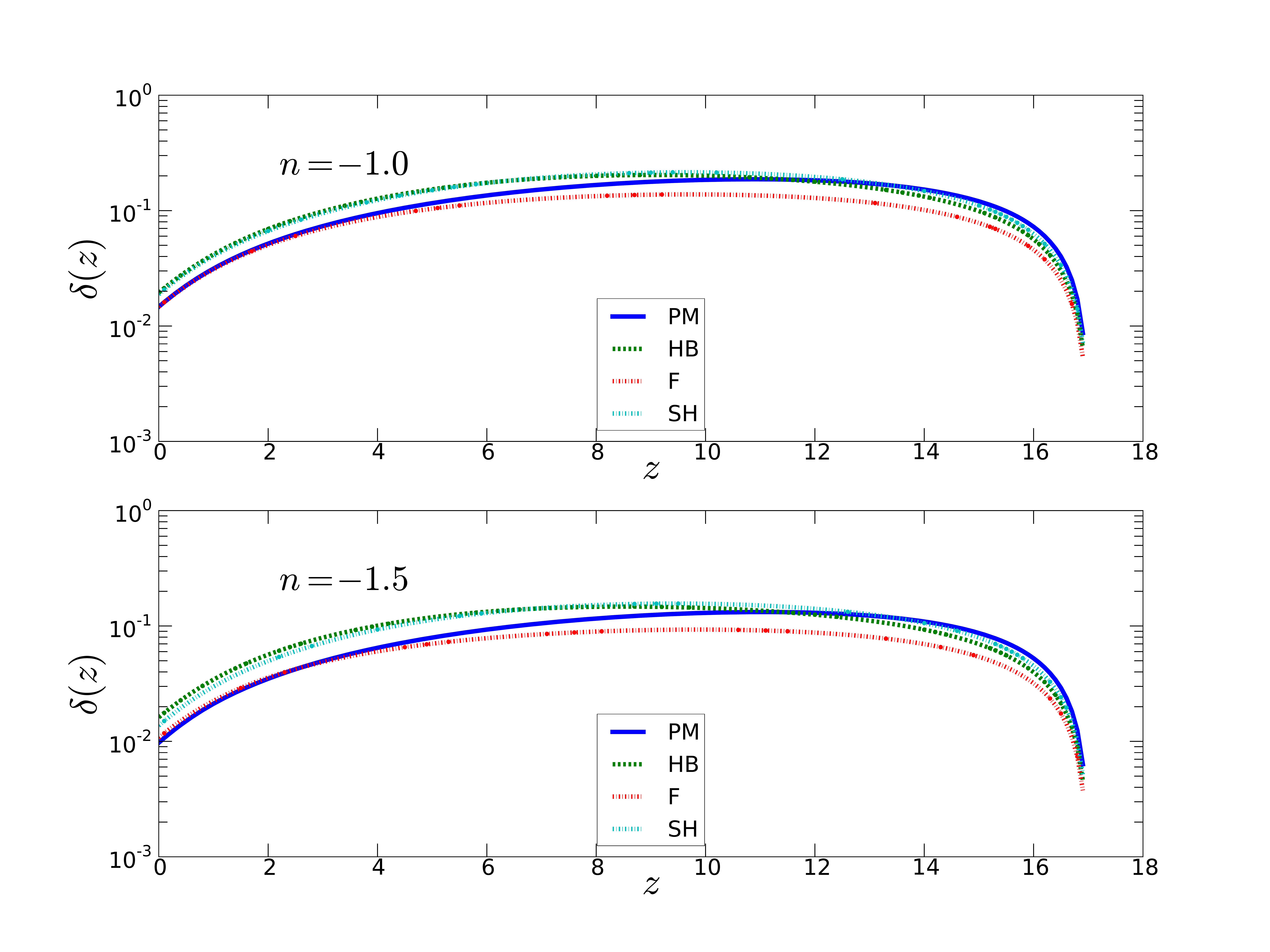}
\caption{The quasar duty-cycle derived for $n=-1.0$ and $n=-1.5$.}\label{fig:Duty}
\end{figure}
\end{center}

On the other hand, the quasar duty-cycle, $\delta(z)$, associated to the SMBHs can be derived from the ratio of $\tau_{\rm DC}(z)$
to the Hubble time. This result is presented in Figure \ref{fig:Duty}. See that $\delta(z)$
is also defined as the fraction of active black holes to their total number. As pointed by \cite{wang06b}, this parameter
is a key to understand how many times and how many black holes are triggered during their lifetimes. In Table \ref{tblduty}, we
show the redshifts where $\delta(z)$ peaks for each CSFR studied here. Note that typically, $\delta$ is maximum in the redshift
range $8.5-11.0$ and this is very curious because the redshift of reionization is $10.5\pm 1.2$ (\citealp{j2011}).

\section{Discussion}\label{sec:4}

In this Letter we present a model to obtain the mass density of SMBHs from the cosmic star formation rate.
The key point to do that is to consider Soltan's argument and that the accretion process can be described by a probabilistic function
directly regulated by the CSFR. Our model permits to determine the function $\rho_{\rm BH}(z)$, the mean radiative efficiency
associated with the growth of the black holes, $\bar\eta(z)$, and the quasar duty-cycle $\delta(z)$. In the literature, the
common way to obtain the mass density of SMBHs is by integration of the QLF as presented in Eq. (\ref{sbrho}) and
considering $\bar{\eta}$ as a constant. However, here we present a different scenario. In particular, we derive $\rho_{\rm BH}(z)$
from the CSFR and then we use the quasar luminosity density in order to obtain the mean radiative efficiency as a function of the
redshift. From $\bar\eta(z)$ is straightforward to obtain the dimensionless mass accretion rate and the quasar duty-cycle. 

If we consider $\rho^{0}_{\rm BH}= (5.9h^{3})\times 10^{5}{\rm M}_{\odot}{\rm Mpc}^{-3}$ \citep{graham07,vika06}, then our model
returns, using the CSFR of \cite{pereira10}, $\rho^{\rm seed}_{\rm BH}=4.60\times 10^{3}{\rm M}_{\odot}{\rm Mpc}^{-3}$ (with $n=-1.0$).
This result is compatible with black hole seeds $\sim 10^{3}\, {\rm M}_{\odot}$ at $z\sim 20$. We also note that the quasar
duty-cycle has a maximum value close to $z\sim 8.5-11$ and so within the observational uncertainties associated to the redshift
of reionization. As the main component of our model is the CSFR, this scenario offers several future possibility of investigations.
One of them is related to the form of the probabilistic function which permits to determine the growth of SMBHs from the CSFR.
Insofar as both star formation and growth of SMBHs can be described in the same way, the predictions of
different probabilistic functions, $P(t_{\rm d})$, can be confronted to different observables at high redshifts. Finally, our
results also show that is not necessary a supercritical Eddington accretion regime to produce the growth of the black hole seeds.

\section*{Acknowledgments}
ESP would like to thank the Brazilian Agency CAPES for support. ODM would like to thank the Brazilian agency CNPq for partial support (grant
300713/2009-6).

\label{lastpage}

\end{document}